\documentclass[conference,10pt,letterpaper]{IEEEtran}

\usepackage[utf8]{inputenc}
\usepackage[T1]{fontenc}
\usepackage{graphicx}
\usepackage[x11names, rgb]{xcolor}
\usepackage{hyperref}
\usepackage{tabularx}
\usepackage{csvsimple}
\usepackage{wrapfig}
\usepackage{amsmath}
\usepackage{amssymb}
\usepackage{array}
\usepackage{multirow}
\usepackage{multicol}
\usepackage{lmodern}
\usepackage{booktabs}
\usepackage{pifont}
\usepackage{tikz}
\usepackage{color}
\usepackage{float}
\usepackage{colortbl}
\usepackage{ltl}
\usepackage{xspace}
\usepackage[capitalise,nameinlink]{cleveref}
\usepackage{varwidth}
\usepackage{stmaryrd}
\usepackage{etex}
\usepackage{listings}
\usepackage{amsmath}
\usepackage{pgfplots}
\usepackage{pgfplotstable}

\usetikzlibrary{calc}
\usetikzlibrary{snakes}
\usetikzlibrary{automata}
\usetikzlibrary{backgrounds}
\usetikzlibrary{decorations.pathreplacing}
\usetikzlibrary{shapes,arrows}
\usetikzlibrary{positioning}
\usetikzlibrary{shadows}

\tikzset{
  joint/.style={line width=0pt,fill,circle,inner sep=0.7pt},
  edgelabel/.style={draw=yellow!50!gray,inner sep=1pt,very thin,text=black,fill=yellow!30,font=\tiny\ttfamily,text height=0.3em,text depth=0.05em,minimum height=0.6em,rounded corners=1},
  ioport/.style={draw=orange!50!gray,inner sep=1pt,very thin,text=black,fill=orange!50,font=\tiny\ttfamily,text height=0.3em,text depth=0.05em,minimum height=0.6em,rounded corners=1},
  component/.style={fill=blue!35,rounded corners=1},
  memory/.style={fill=gray!45,rounded corners=1},
  initialb/.style={fill=gray!45,rounded corners=1},
  cell/.style={inner sep=2pt,rounded corners=1,fill=gray!80!black}
}

\newcommand{\dtime}{\mbox{\ensuremath{\mathcal{T}\hspace{-2.5pt}\scalebox{0.8}{\ensuremath{\boldsymbol{i}\boldsymbol{m}\boldsymbol{e}}}}}}
\newcommand{\from}{\ensuremath{\colon}}

\renewcommand{\to}{\ensuremath{\rightarrow}}

\newcommand{\cells}{\ensuremath{\mathbb{C}}\xspace}

\newcommand{\branch}[2]{\ensuremath{#1 \! \wr \hspace{-0.6pt} #2}}
\newcommand{\term}{\ensuremath{\tau}}
\newcommand{\fterm}{\ensuremath{\term_{F}}\xspace}
\newcommand{\pterm}{\ensuremath{\term_{P}}\xspace}

\newcommand{\name}[1]{{\text{\texttt{#1}}}}
\newcommand{\const}[1]{\ensuremath{\text{\texttt{#1}}()}}
\newcommand{\constidx}[2]{\ensuremath{\text{\texttt{#1}}_{#2}()}}
\newcommand{\terms}{\ensuremath{\mathcal{T}}\xspace}

\newcommand{\sep}{\ensuremath{\quad | \quad}}
\newcommand{\upd}[2]{\ensuremath{\llbracket \hspace{1pt} #1 \! \leftarrowtail #2 \hspace{1pt} \rrbracket}}
\newcommand{\sats}{\ensuremath{\;\vDash_{\!\langle \hspace{-1pt} \cdot \hspace{-1pt} \rangle}}}

\newcommand{\pterms}{\ensuremath{\terms_{\!P\hspace{-0.5pt}}}\xspace}
\newcommand{\fterms}{\ensuremath{\terms_{\!F\hspace{-0.5pt}}}\xspace}

\newcommand{\functions}{\ensuremath{\mathcal{F}}\xspace}

\newcommand{\assign}[1]{\ensuremath{\langle #1 \rangle}}

\newcommand{\inames}{\ensuremath{\mathbb{I}}\xspace}
\newcommand{\onames}{\ensuremath{\mathbb{O}}\xspace}
\newcommand{\pnames}{\ensuremath{\mathbb{P}}\xspace}
\newcommand{\fnames}{\ensuremath{\mathbb{F}}\xspace}

\lstdefinestyle{tsl}{
  language=c,
  basicstyle=\footnotesize\ttfamily,breaklines=true,
  mathescape=false,
  frame=none,
  literate=
    {->}{{{\color{blue!80}\ \ \texttt{->}\;\ }}}1
    {<-}{{{\color{red!50!black}\ \ \texttt{<-}\;\ }}}1
    {<->}{{{\color{blue!80}\ \ \texttt{<->}\;\ }}}1
    {\&\&}{{{\color{blue!80}\ \ \texttt{\&\&}\;\ }}}1
    {||}{{{\color{blue!80}\ \ \texttt{||}\;\ }}}1
    {!}{{{\color{blue!80}\ \ \texttt{!}\;\ }}}1
    {[}{{{\color{red!50!black}\ \ \texttt{[}\;\ }}}1
    {]}{{{\color{red!50!black}\ \ \texttt{]}\;\ }}}1
    {\{}{{{\color{black}\ \ \texttt{\{}\;\ }}}1
    {\}}{{{\color{black}\ \ \texttt{\}}\;\ }}}1
    {()}{{\ \ \texttt{()}\;\ }}1
    {;}{{\ \ \texttt{;}\;\ }}1,    
  deletekeywords={Double,Int,init,Bool},
  identifierstyle={\ttfamily\color{black}},
  keywordstyle=[1]{\ttfamily\color{green!60!black}},
  keywordstyle=[2]{\ttfamily\color{violet}},
  keywordstyle=[3]{\ttfamily\color{blue!80}},
  keywordstyle=[4]{\ttfamily\color{yellow!50!black}},
  keywordstyle=[5]{\ttfamily\color{red}},
  keywordstyle=[6]{\ttfamily\color{cyan!40!black}},
  morekeywords=[1]{COUNTUP, COUNTDOWN, INCMIN, INCSEC, IDLE, ZERO, RESET, COUNTING, ANYKEY, START, STARTANDMIN, STARTANDSEC},
  morekeywords=[2]{xor, press, tillAnyInput},
  morekeywords=[3]{X, W},
  morekeywords=[4]{x, y},
  morekeywords=[6]{always, guarantee, initially},
  commentstyle=\color{orange!50!black}, 
  stringstyle=\color{red!50!black},
}

\IEEEoverridecommandlockouts

\begin{document}

\title{Syntroids: Synthesizing a Game for FPGAs \\ using Temporal Logic
  Specifications}

\author{
  \IEEEauthorblockN{
    Gideon Geier,
    Philippe Heim,
    Felix Klein and
    Bernd Finkbeiner
  } 
  \IEEEauthorblockA{
    Reactive Systems Group, Saarland University, Germany \\
    \texttt{\{geier,heim,klein,finkbeiner\}@react.uni-saarland.de}
  }
  \thanks{
    Supported by the European Research Council (ERC) Grant \mbox{OSARES}
    (No.\ 683300) and the German Research Foundation (DFG) as part of the
    Collaborative Research Center Foundations of Perspicuous Software
    Systems (TRR 248, 389792660).
  }
}

\maketitle

\newcommand{\asgn}{\ensuremath{\!\!\!\!:=\!\!\!\!}}
\newcommand{\until}{\LTLuntil}
\newcommand{\wuntil}{\LTLweakuntil}
\newcommand{\release}{\LTLrelease}
\newcommand{\finally}{\LTLfinally}
\newcommand{\globally}{\LTLglobally}
\newcommand{\func}[1]{\text{\textsf{#1}}}
\newcommand{\un}[2]{\upd{\name{#1}}{#2}}
\newenvironment{codeblockeq}{\par \vspace{-0.7em} \begin{small}\begin{equation*}}{\end{equation*}\end{small}}
\newenvironment{codeblock}{\par \vspace{-0.7em}\begin{small}\begin{equation*}\begin{array}{rcl}}{\end{array}\end{equation*}\end{small}}

\newcommand{\clash}{\textsc{C}$\hspace{-0.5pt}\lambda\hspace{-0.5pt}$\textsc{aSH}}

\definecolor{bestcolor}{RGB}{150,150,255}
\definecolor{failcolor}{RGB}{255,0,0}
\newcommand{\us}[1]{\cellcolor{blue!20}#1}
\newcommand{\bs}[1]{\textbf{#1}}
\newcommand{\usb}[1]{\us{\bs{#1}}}
\newcommand{\fail}{\color{red!70!black}{\!$>99999$\!}}

\newcommand{\tableData}{
\begin{table*}
\renewcommand{\arraystretch}{1}
\caption{}
\label{table_results}
\centering
\begin{tabularx}{\linewidth}{p{114pt}|cc|cc|c|ccc|ccc|ccc}
    \bfseries Module & \multicolumn{2}{c|}{\bfseries G} & \multicolumn{2}{c|}{\bfseries A} & \bfseries M & \multicolumn{3}{c|}{\bfseries Bosy} & \multicolumn{3}{c|}{\bfseries Bowser} & \multicolumn{3}{c}{\bfseries Strix} \\
                     & L  & T  & L & T &  & Time & Lat  & Gat & Time & Lat & Gat & Time & Lat  & Gat
\\ \hline
ActionConverter (AC)        & 4  & 0  & 0 & 0   & 1           & \bs{0.316}   & 1      & 8          & \us{4.384}  & \usb{0}   & \usb{4}    & 1.192         & \bs{0}     & \bs{4}   \\
Cockpitboard (CB)           & 12 & 0  & 0 & 0   & 1           & 1548.48      & 1      & 11         & \us{227.136}& \usb{0}   & \usb{7}    & \bs{6.512}    & \bs{0}     & \bs{7}   \\
EnemyModule (EM)            & 4  & 0  & 0 & 0   & 1           & \bs{0.272}   & 1      & 6          & \us{0.904}  & \usb{0}   & \usb{2}    &1.196          & \bs{0}     & \bs{2}   \\
Gamelogic (GL)              & 15 & 6  & 0 & 0   & \!$3 \lor 4$\!      & \fail        & -      & -          & 13230.5     & \bs{2}    & 226        & \!\usb{696.288}\! & \usb{2}    & \usb{29} \\
GamemodeChooser (GC)        & 7  & 0  & 4 & 0   & 1           & 111.072      & 1      & 100        & 13247.0     & \bs 0     & 2377       & \usb{2.164}   & \us 1      & \usb{35} \\
Gamemodule (GM)             & 3  & 0  & 3 & 0   & 1           & \bs{0.328}   & 1      & 10         & \us{1.056}  & \usb 0    & \usb 3     & 1.288         & 1          & 11       \\
LedMatrix (LM)              & 14 & 13 & 0 & 1   & $\leq 32$   & \fail        & -      & -          & \fail       & -         & -          & \!\usb{53732.0}\! & \usb 5     & \!\usb{101}\!\\
Radarboard (RB)             & 13 & 0  & 0 & 0   & 1           & 41319.2      & 1      & 10         & \usb{26.448}& \usb 0    & \usb 6     & 79.376        & \bs 0      & \bs 6    \\
RegisterManager (RM)        & 5  & 0  & 0 & 0   & 1           & 0.292        & 1      & 4          & \usb{0.164} & \usb{0}   & \usb{0}    & 1.084         & \bs{0}     & \bs{0}   \\
RotationCalculator (RC)     & 5  & 0  & 3 & 0   & 1           & \bs{1.324}   & 1      & 18         & \us{8045.37}& \usb 0    & \usb 9     & 1.66          & 1          & 22       \\
SPI (SPI)                   & 6  & 9  & 2 & 0   & 3           & \fail        & -      & -          & 13214.7     & \bs 3     & 413        & \usb{3.608}   & \usb 3     & \usb{72} \\
SPIReadClk (SPI$_{\text{R}}$)          & 2  & 0  & 0 & 0   & 1           & \bs{0.272}   & 1      & 4          & \us{0.348}  & \usb 0    & \usb 2     & 1.168         & \bs 0      & \bs 2    \\
SPIReadManag (SPI$_{\text{R}} $)        & 9  & 2  & 1 & 1   & 2           & 3497.26      & 1      & 31         & \us{11821.0}& \usb 1    & \usb{10}   & \bs{14.684}   & 2          & 27       \\
SPIReadSdi (SPI$_{\text{R}}$)          & 2  & 0  & 2 & 0   & 1           & \bs{0.304}   & 1      & 5          & \us{0.844}  & \usb 0    & \usb 1     & 1.36          & 1          & 5        \\
SPIWriteClk (SPI$_{\text{W}}$)         & 2  & 0  & 0 & 0   & 1           & \bs{0.276}   & 1      & 6          & \us{3.628}  & \usb 0    & \usb 4     & 1.196         & \bs 0      & \bs 4    \\
SPIWriteManag (SPI$_{\text{W}}$)       & 7  & 2  & 3 & 1   & 2           & 196.808      & \bs 1   & \bs 6          & 61.704      & \bs 1     & \bs 6      & \usb{2.22}    & \usb 1     & \usb 6   \\
SPIWriteSdi (SPI$_{\text{W}}$)         & 3  & 0  & 5 & 0   & 1           & \bs{0.396}   & 1      & 15         & \us{12.536} & \usb 0    & \usb 4     & 1.3           & 1          & 11       \\
Scoreboard (SB)             & 7  & 0  & 0 & 0   & 1           & \bs{1.26}    & 1      & 8          & \us{15.576} & \usb 0    & \usb 4     & 1.516         & \bs 0      & \bs 4    \\
Sensor (Sen)                & 2  & 4  & 0 & 4   & 4           & \us{7429.74} & \usb 2 & \usb{29}   & \fail       & -         & -          & \bs{1.912}    &  4         & 70  \\
SensorInit (Sen)            & 2  & 12 & 0 & 0   & 9           & 159.076      & \bs 4  & 95         & 6613.8      & \bs 4     & 84         & \usb{3.676}   & \usb 4     & \usb{46} \\
SensorPart (Sen)            & 9  & 9  & 0 & 0   & 8           & 1985.21      & \bs 3  & 34         & 12224.9     & \bs 3     & 64         & \usb{13.864}  & \usb 3     & \usb{30} \\
SensorRegister (RM)         & 1  & 0  & 0 & 0   & 1           & 0.292        & 1      & 2          & \usb{0.048} & \usb 0    & \usb 0     & 1.188         & \bs 0      & \bs 0    \\
SensorSelector (SS)         & 5  & 0  & 4 & 0   & 1           & \fail        & -      & -          & \usb{37.884}& \usb 0    & \usb 0     & 277.288       & 1          & 17       \\
SensorSubmodulChooser (Sen) & 1  & 4  & 6 & 0   & 4           & \us{766.084} & \usb 2 & \us{44}    & 13007.8     & \bs 2     & 369        & \bs{3.176}    & 3          & \bs{39}       
\end{tabularx}
\end{table*}
}

\newcommand{\tableSizesAll}{
\begin{table}
\renewcommand{\arraystretch}{1.1}
\begin{tabularx}{\linewidth}{X|l|rr}
Module(s) & T & L & D \\ \hline
EnemyModule                      & BW & 4104 & + 46 \\
Gamelogic                        & ST & 4085 & + 27 \\
Radarboard                       & BW & 3946 & -112 \\
Cockpitboard                     & BW & 4057 & -  1 \\
Scoreboard                       & BW & 4066 & +  8 \\
Gamemodule                       & BW & 4041 & - 17 \\
ActionConverter                  & BW & 4069 & + 11 \\
GamemodeChooser                  & ST & 4104 & + 46 \\
RotationCalculator               & BW & 3919 & -139 \\
LedMatrix                        & ST & 4235 & +177 \\
RegisterManager + SensorRegister & BW & 4057 & -  1 \\
Sensor:                          &    & 4425 & +367 \\
- Sensor                         & BS \\ 
- SensorInit                     & ST \\
- SensorPart                     & ST \\
- SensorSubmodulChooser          & BS \\
- SensorSelector                 & BW \\       
SPI                              & ST & 4223 & +165 \\
SPIRead:                         &    & 4232 & +174 \\
- SPIReadManag                   & BW \\
- SPIReadClk                     & BW \\
- SPIReadSdi                     & BW \\                  
SPIWrite:                        &    & 4316 & +258 \\
- SPIWriteManag                  & ST \\
- SPIWriteClk                    & BW \\
- SPIWriteSdi                    & BW \\                 
\end{tabularx}
\end{table}
}

\newcommand{\tableSizesCoarse}{
\begin{table}
\renewcommand{\arraystretch}{1.1}
\begin{tabularx}{\linewidth}{l|X|r}
N & M & D \\ \hline
All             & All & +642 \\ \hline
Logic           & EnemyModule, Gamelogic & +168\\ \hline
Part-Logic      & EnemyModule Gamelogic Radarboard Cockpitboard Scoreboard Gamemodule & +37 \\ \hline
Converter       & ActionConverter GamemodeChooser RotationCalculator & -6 \\ \hline
Logic           & EnemyModule Gamelogic Radarboard Cockpitboard Scoreboard ActionConverter GamemodeChooser RotationCalculator Gamemodule & -62 \\ \hline
Sensor-Logic    & Sensor SensorRegister RegisterManager & +315 \\ \hline  
SPI             & SPI SPI-Read SPI-Write & +535 \\ \hline
Sensor          & Sensor SensorRegister RegisterManager SPI SPI-Read SPI-Write \textbf{todo} Write them out? & +744 \\ \hline 
IO              & Sensor SensorRegister RegisterManager SPI SPI-Read SPI-Write LedMatrix \textbf{todo} Write them out & +822
\end{tabularx}
\end{table}
}

\newcommand{\diagram}{
\pgfplotstableread[col sep=comma]{sizes2.csv}{\tablesizes}
\begin{figure}[h]
\scalebox{1}{
\begin{tikzpicture}
    \begin{axis}[
        width=\axisdefaultwidth*1.1,
        height=\axisdefaultheight,
        ymin=2000,
        ymax=8000,
        ybar,
        width=\linewidth,
        height=150pt,
        xtick=data,
        xticklabels from table={\tablesizes}{M},
        xticklabel style={yshift=3pt},
        enlarge x limits={abs=0.6},
        axis y line*=left,
        legend style={at={(0.9,0.98)}, anchor=north}
    ]
            \addplot table [
                x expr=\coordindex,
                y index=3,
                col sep=comma,
            ] {\tablesizes};
        \addplot[orange,sharp plot,update limits=false] coordinates {(-1,4058) (16,4058)} node[above] at (axis cs:3.5,4058) {\footnotesize handmade LCs};
        \addplot[red,sharp plot,update limits=false] coordinates {(-1,7680) (16,7680)} node[below] at (axis cs:3,7680) {\footnotesize max LCs};
        \legend{LCs}
    \end{axis}

    \begin{axis}[
        width=1.1*\axisdefaultwidth,
        height=\axisdefaultheight,
        ymin=10,
        ymax=30,
        axis y line*=right,
        axis x line=none,
        width=\linewidth,
        height=150pt,
        enlarge x limits={abs=0.6},
        legend style={at={(0.68,0.98)}, anchor=north},
    ]
            \addplot table [
                only marks,
                x expr=\coordindex,
                y index=5,
                col sep=comma,
            ] {\tablesizes};
        \addplot[purple,update limits=false] coordinates {(-1,22.41) (16,22.41)} node[above] at (axis cs:2,22.41) {\footnotesize handmade MHz};
        \legend{MHz}
    \end{axis}

\end{tikzpicture}
}\vspace{-22pt}
    \caption{LCs \& timing for multiple modules swapped. \mbox{\ }}
    \label{fig:diagram}
\end{figure}
}

\newcommand{\diagramall}{
\pgfplotstableread[col sep=comma]{sizes.csv}{\loadedtablesec}
\begin{figure}[b]
  \vspace{-1.5em}
  \scalebox{1}{

\begin{tikzpicture}
    \begin{axis}[
        width=1.1*\axisdefaultwidth,
        height=\axisdefaultheight,
        ymin=2000,
        ymax=8000,
        ybar,
        axis y line*=left,
        width=\linewidth,
        height=150pt,
        xtick=data,
        xticklabels from table={\loadedtablesec}{Module(s)},
        xticklabel style={rotate=-35,anchor=north west,yshift=2pt,xshift=-2pt},
        enlarge x limits={abs=0.6},
        legend style={at={(0.9,0.98)}, anchor=north}
    ]
            \addplot table [
                x expr=\coordindex,
                y index=2,
                col sep=comma,
            ] {\loadedtablesec};

        \addplot[orange,sharp plot,update limits=false] coordinates {(-1,4058) (16,4058)} node[above] at (axis cs:5,4058) {\footnotesize handmade LCs};
        \addplot[red,sharp plot,update limits=false] coordinates {(-1,7680) (16,7680)} node[below] at (axis cs:5,7680) {\footnotesize max LCs};
        \legend{LCs}
    \end{axis}

    \begin{axis}[
        width=1.1*\axisdefaultwidth,
        height=\axisdefaultheight,
        ymin=10,
        ymax=30,
        axis y line*=right,
        axis x line=none,
        width=\linewidth,
        height=150pt,
        enlarge x limits={abs=0.6},
        legend style={at={(0.68,0.98)}, anchor=north},
    ]
            \addplot table [
                only marks,
                x expr=\coordindex,
                y index=5,
                col sep=comma,
            ] {\loadedtablesec};
        \addplot[purple,update limits=false] coordinates {(-1,22.41) (16,22.41)} node[above] at (axis cs:2,22.41) {\footnotesize handmade MHz};
        \legend{MHz}
    \end{axis}
\end{tikzpicture}}
\vspace{-28pt}
\caption{LCs \& timing for a single module swapped. \mbox{\ }}
\label{fig:diagramall}
\end{figure}
}


\input{architecture.tex}

\begin{abstract}
  We present \textit{Syntroids}, a case study for the automatic
  synthesis of hardware from a temporal logic
  specification. \textit{Syntroids} is a space shooter arcade game
  realized on an FPGA, where the control flow architecture has been
  completely specified in Temporal Stream Logic (TSL) and implemented
  using reactive synthesis.  TSL is a recently introduced temporal
  logic that separates control and data.  This leads to scalable
  synthesis, because the cost of the synthesis process is independent
  of the complexity of the handled data.
  
  In this case study, we report on our experience with the TSL-based
  development of the \textit{Syntroids} game and on the implementation
  quality obtained with synthesis in comparison to manual programming.
  We also discuss solved and open challenges with respect to currently
  available synthesis tools.
\end{abstract}

\tikzset{
  specified/.style={rectangle},
  unspecified/.style={ellipse},  
}

\section{Introduction}

\begin{figure*}[t]
  \centering
  \begin{tikzpicture}
    \node at (0,0) {\includegraphics[scale=0.045]{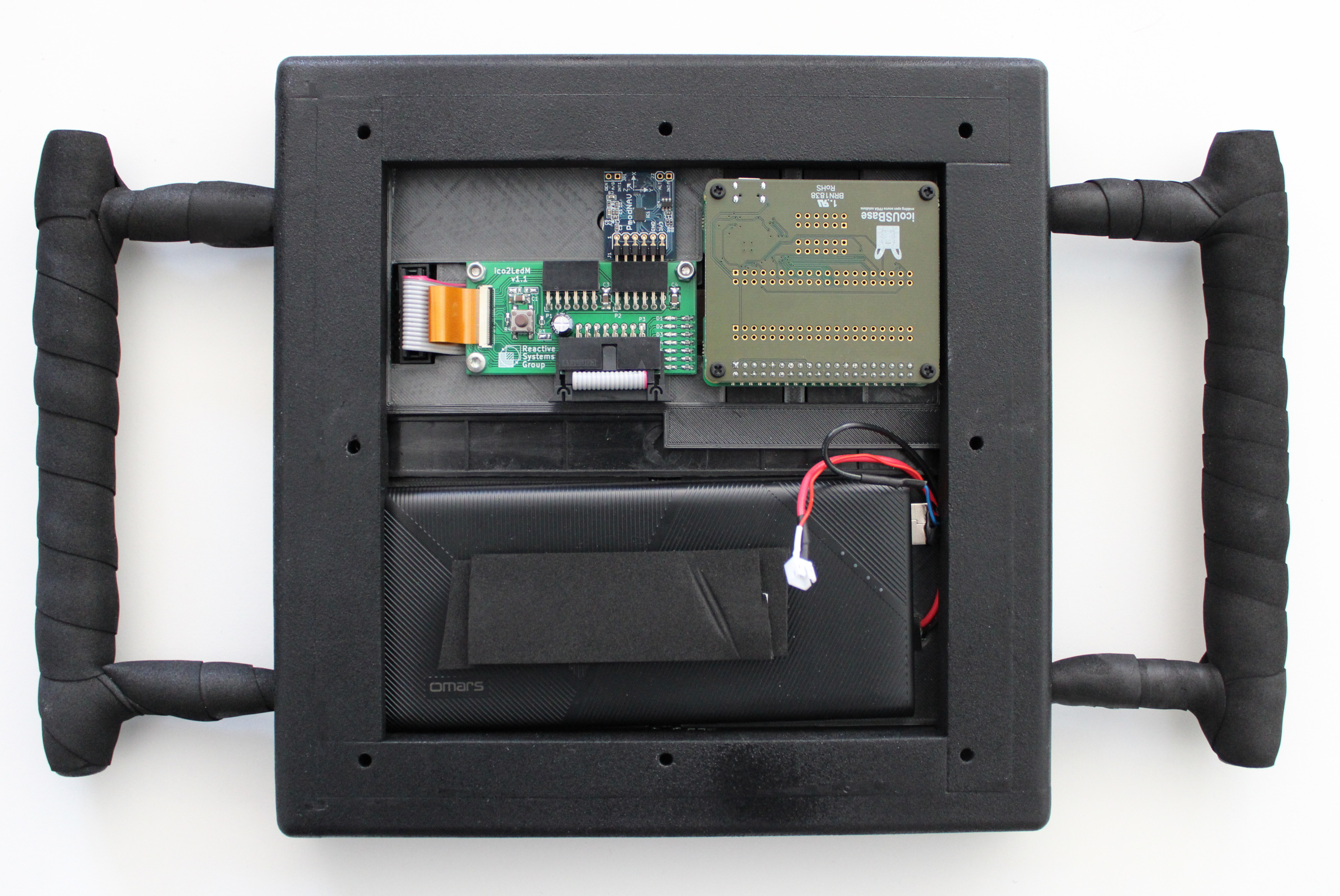}};
    \node at (6.1,0) {\includegraphics[scale=0.045]{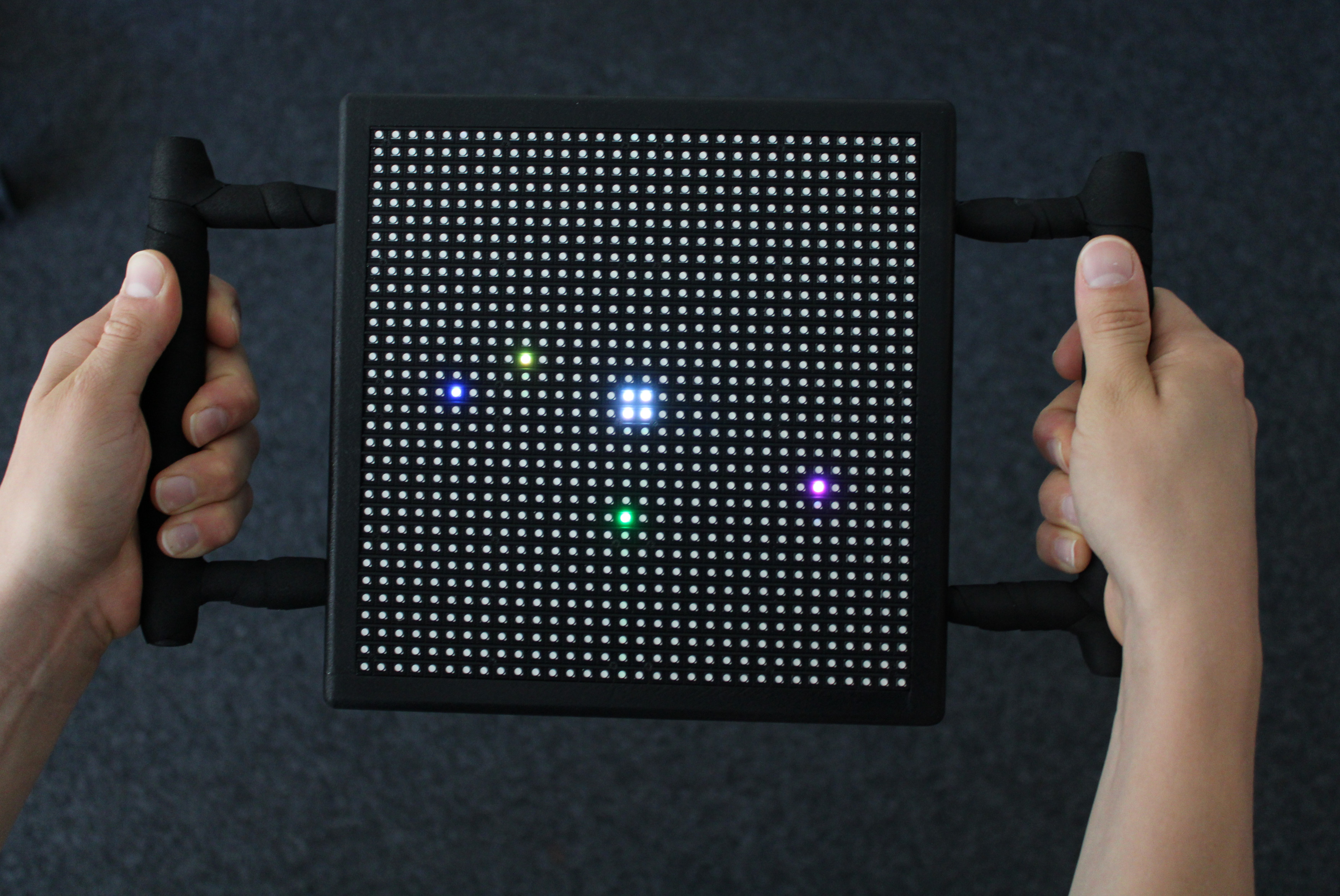}};
    \node at (12.2,0) {\includegraphics[scale=0.045]{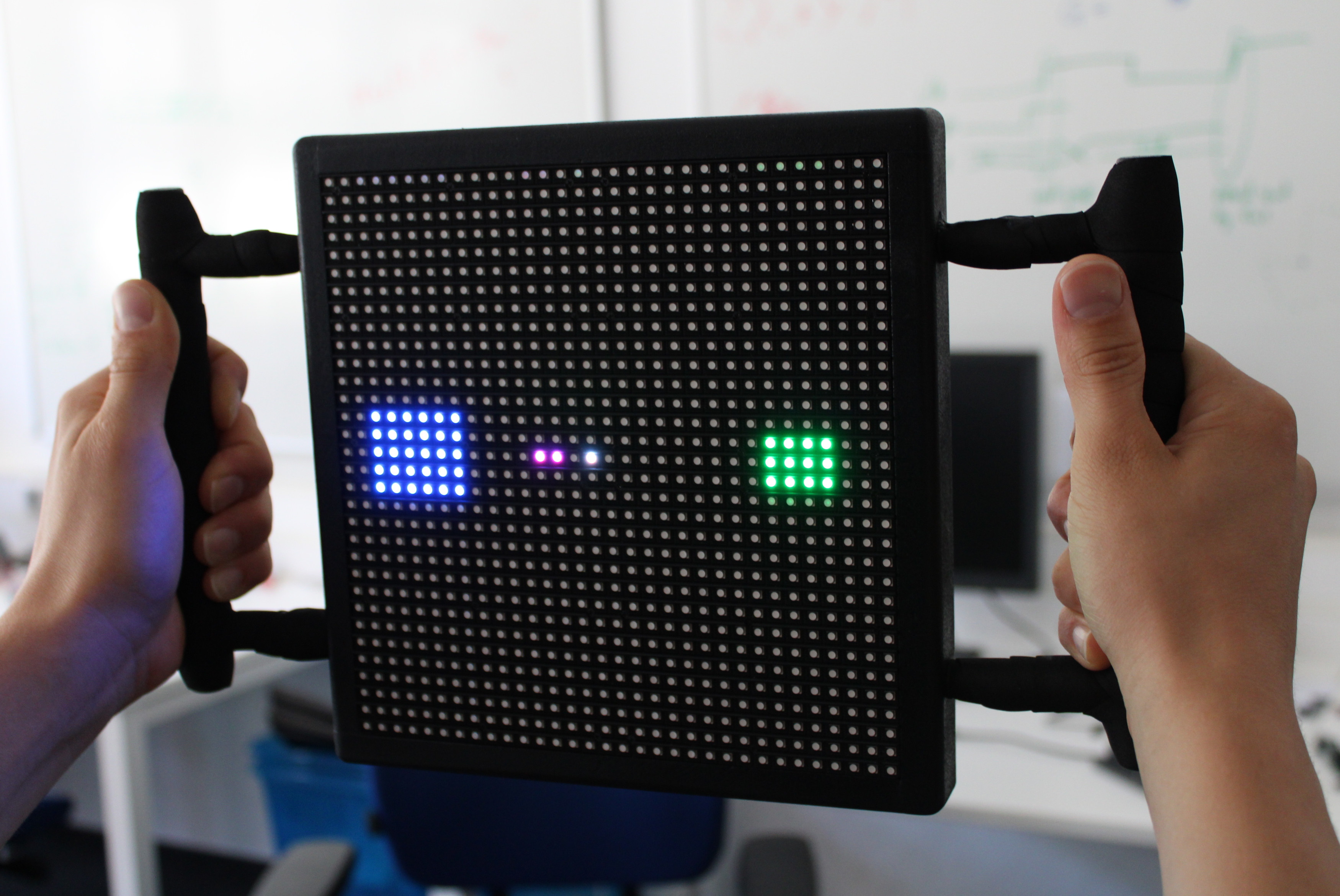}};
  \end{tikzpicture}
  \vspace{-1em}
  \caption{The hardware (left), radar mode (center) revealing four enemies, and cockpit mode (right) showing three enemies facing the player.}
  \label{fig:game}
\end{figure*}

\noindent
Computationally controlled systems that are embedded into physical products
are of ever-growing importance in modern life. They range from simple devices, like a kitchen timer or a heating controller, to enormously complex ones like
autonomous vehicles and aircraft. For safety-critical systems, the standard
design flow is to first manually write an implementation and then verify the implementation against a formal specification.

An attractive alternative to this design flow is offered by \emph{reactive synthesis},
which automatically creates a correct-by-construction implementation from
a specification given in a temporal logic. In practice, however, applying the currently available synthesis tools is difficult. Even though there has been some success in synthesizing
\emph{control-intensive} systems, such as the AMBA arbiter~\cite{amba},
synthesis tools often fail due to the complexity of the handled \emph{data}.  For standard reactive synthesis, all data structures must be encoded on the bit-level, which, for complex data, results in a far too large state space.

Recently, a new temporal logic, \emph{Temporal Stream Logic} (TSL), has
been introduced that specifically addresses this problem~\cite{tsl}. TSL
separates the specification of the control structure from the data
transformations.  This leads to scalable synthesis, because the cost
of the synthesis process is independent of the complexity of the
handled data.

In this paper, we describe a case study in which we apply 
TSL-based synthesis to the development of a non-trivial
arcade game. 
We develop \textit{Syntroids}, a space shooter
game realized on an FPGA. The design of the game involves several
data-intensive features
that need to be handled by TSL, like reading data
from an external sensor using an SPI interface, displaying data on a
multi-color LED matrix, and managing an open number of enemies in the
game's world. To the best of our knowledge, this is the most complex case
study for reactive synthesis to date.

The \textit{Syntroids} game is  controlled via the
orientation and movement of a physical screen, similar to modern
smartphone games. The player is inside a spaceship and has to shoot
asteroids, also referred to as enemies, rushing at the spaceship from
all directions.  If the spaceship is hit by an asteroid, then the game
is over.  The player gets a point for every asteroid taken down.  As
the game progresses, the difficulty increases with new enemies
moving faster and faster.  At all times, the player can switch back and forth
between three different game modes:
\begin{enumerate}

\item If the device is held horizontally (screen is upward) the game switches to
  \textit{radar mode}. The radar mode shows a top-view of the
  environment, in which the player can easily determine where and how
  close enemies are.

\item If the device is held vertically (screen towards the player), the game switches to
  \textit{cockpit mode}.  Cockpit mode shows the view through the
  windshield of the spaceship.  To look into different directions, the
  player has to turn accordingly.  Once enemies are close enough to be visible,   the player can shoot at them by aiming the device straight at one of the enemies and pushing the device quickly into the direction of the enemy and back again.  The spaceship's laser gun then
  instantly destroys the asteroid.

\item The \textit{score mode} shows the score, depicted as one dot per
  point. This mode is chosen by holding the device upside down over the
  player's head. The score is shown in the color of the
  most dangerous asteroid.

\end{enumerate}
When the player's spaceship is hit by an asteroid, a game over screen is shown and the game can be
restarted by doing a shooting gesture. Pictures of the device's
hardware, as well as radar and cockpit mode are shown in \cref{fig:game}.

\section{Game Architecture}

\noindent The game's architecture, depicted in~\cref{fig:architecture}, can be organized according to four tasks:
management of the game logic, generating LED matrix output, SPI in- and output handling, and
controlling sensor data acquirement, as well as storing and converting
them into usable signals.

\subsubsection{Gameplay}
The game logic is implemented by the \textit{GameLogic} module and the \textit{EnemyModules}.
Every \textit{EnemyModule} stores the data of a single enemy and moves or
resets the enemy to a new position, represented as a polar
coordinate.
Each enemy gets an individual move-clock-signal from 
\textit{GameLogic} indicating if it must move.

The \textit{GameLogic} also generates random starting angles as respawn points for the enemies.
Therefore, it checks, whether an enemy was shot or the player is hit to generate game over or reset signals.
Furthermore it handles the score, which increases with every enemy taken down and resets if the game restarts.
Overall, the \textit{GameLogic} module can handle a variable number of enemies.

\subsubsection{Video Output}
The \textit{GameModule} manages the drawing process of the game.
The three drawing modules (\textit{Cockpit-}, \textit{Radar-} and \textit{ScoreBoard}) work in parallel.
They iterate over all pixels, where they output data for one pixel every clock cycle.
Based on the game mode and whether the game is over
the \textit{GameModule} chooses the right pixel to be written to the video-memory of the \textit{LedMatrix} module. 

The \textit{ScoreBoard} manages the printing of the score and the game over screen, depending on the state of the game.
The \textit{RadarBoard} prints the radar by calculating Cartesian coordinates 
for every enemy periodically with respect to the screen's orientation.
The cockpit mode is managed by the \textit{CockpitBoard}, whose main
task is drawing the enemies as squares depending on their distance and
rotation, as well as on the orientation of the player. Therefore, the module
checks for every displayed pixel, whether and which enemy it displays.

The LED matrix screen is controlled by the \textit{LedMatrix} module
also providing the video memory. Furthermore, it allows to change pixels
independently of the update process of the physical LED matrix.

\subsubsection{User input}
A sensor device, attached via PMod interface, is used to determine
orientation and movement of the device.  To this end, the
accelerometer and gyroscope embedded as part of the sensor are used.
The communication with the sensor works via a 4-pin SPI.

The \textit{SPI} module implements standard SPI communication
by coordinating the \textit{SPI Write} and \textit{SPI Read} modules, 
which are split into several submodules.
For reading and writing, respectively, one of the submodules manages a state, while the others use it to generate output.
The \textit{SPI Read} module also has to retrieve data from the \name{sdo} input pin.
The modules all work for variable serial clock speeds.

The sensor submodules are coordinated by the \textit{Sensor} module. 
In the beginning the \textit{Sensor} module selects the \textit{SensorInit} module, which initializes 
the device.
Afterwards the \textit{SensorPart} modules are scheduled to read their assigned 
registers, e.g., all accelerometer registers.
They communicate with the \textit{SPI} module, the \textit{RegisterManager} and the \textit{SensorSelector}, where
the \textit{SubmoduleChooser} only forwards the signals of the currently selected module.
The device uses chip select pins for accessing the different sensors. The \textit{SensorSelector} 
integrates these with the SPI communication and selects them according to
which part of the sensor the module is currently communicating with.

\subsubsection{Data handling}

\begin{figure*}[t]
  \centering
  \scalebox{0.65}{\usebox{\boxarchitecture}}
  \vspace{-0.8em}
  \caption{The architecture of the game. The blue highlighted
    components have been synthesized from specifications written in
    TSL.}
  \label{fig:architecture}
\end{figure*}

To read the sensor data asynchronously and independent of the current communication state,
the data is stored in separate \textit{SensorRegisters}.
The \textit{RegisterManager} controls all the registers. Moreover since the sensor
values are read in two steps, it caches the intermediate values. 

The conversion of the data to meaningful inputs of the game is handled by 
three  converter modules using several (empirical gained) threshold values for the specific sensor outputs.   
The \textit{RotationCalculator} calculates the player's absolute rotation from his starting position,
using the gyroscope's $ x $- and $ z $-axis, depending on the mode of the game. 
The rotation is gained by integrating the gyroscope's output to the rotation speed.
The \textit{GamemodeChooser} works on the gyroscope's $ y $-axis, which is used to
 differentiate between the three game modes based on the rotation.
The \textit{ActionConverter} recognizes shooting and resetting the game, 
for which the $ z $-axis of the accelerometer is used.
To avoid misdetection due to centrifugal force it also utilizes the three gyroscope values.

\section{Temporal Stream Logic}

\noindent The control flow behavior of all highlighted modules (blue)
in the architecture of \cref{fig:architecture} has been specified
using Temporal Stream Logic (TSL)~\cite{tsl}. The logic introduces a
clean separation between pure data transformations and temporal
control. If a TSL specification is realizable, then it can be turned
into a Control Flow Architecture, an abstract representation of the
hardware architecture that covers all possible behavior switches. In
combination with concretizations for pure data transformations and the
functional hardware description language
{\clash}~\cite{clash2015}, the control flow then is implemenented
on the FPGA.

Temporal Stream Logic builds on the notion of \mbox{\textit{updates}},
such as \mbox{$ \upd{\name{y}}{\name{f}~\name{x}} $} expressing that
on every clock cycle the pure function~\name{f} is applied to the
input stream~\name{x} and the result is piped to the output
stream~$ \name{y} $. These updates are combined 
with predicate evaluations guiding the temporal control flow
decisions. In combination with Boolean and temporal operations, the
logic allows for expressing even complex, temporally evolving
architectures using only a short, but precise description of the
temporal control.

The advantage of TSL is that function and predicate names, as
used by the specification, are only considered as symbolic
literals. The semantics of the logic then guarantee that synthesized systems
satisfy the specified behavior for all possible implementations of
these literals. They are 

\begin{figure}[h]
  \centering
  \vspace{-0.8em}
  \begin{tikzpicture}[scale=0.8]

    \node[anchor=east,inner sep=0pt] at (-2.8,-1.5) {
      \small
      \begin{tabular}{c}
        inputs: \\[0.1em] $ \inames $
      \end{tabular}
    };

    \node at (0,1.48) {
      \small
      cells: $ \cells $
    };

    \node[anchor=west,inner sep=0pt] at (2.8,-1.5) {
      \small
      \begin{tabular}{c}
        outputs: \\[0.1em]
        $ \onames $
      \end{tabular}
    };

    \node at (0,0) {
      \begin{tikzpicture}[xscale=0.8,yscale=0.56]
        \node[fill, fill=blue!35,minimum height=5.5em, minimum width=10.5em] (C) {};

        \node at (C) {
          \small
          \begin{tabular}{c}
            \textit{reactive system} \\[0.4em]
            \textit{implementing a} \\[0.4em]
            \textit{TSL specification~$ \varphi $}
            \end{tabular}
        };

        \node[minimum size=0.9em] (H0) at (0,2.05) {};
        \node[minimum size=0.9em] (H1) at (0,2.7) {};
        \node[minimum size=0.9em] (H2) at (0,3.9) {};

        \path[->,>=stealth,line width=0.7pt]
        ($ (C.west) + (-0.6,-1.3) $) edge ($ (C.west) + (0,-1.3) $)
        ($ (C.west) + (-0.6,-0.6) $) edge ($ (C.west) + (0,-0.6) $)
        ($ (C.west) + (-0.6,-0.3) $) edge ($ (C.west) + (0,-0.3) $)
        ($ (C.east) + (0,-1.3) $) edge ($ (C.east) + (0.6,-1.3) $)
        ($ (C.east) + (0,-0.6) $) edge ($ (C.east) + (0.6,-0.6) $)
        ($ (C.east) + (0,-0.3) $) edge ($ (C.east) + (0.6,-0.3) $)
        ;

        \node at ($ (C.west) + (-0.35,-0.85) $) {\scalebox{0.6}{$ \vdots $}};
        \node at ($ (C.east) + (0.25,-0.85) $) {\scalebox{0.6}{$ \vdots $}};

        \draw[line width=0.7pt,-,>=stealth,gray]
        ($ (C.east) + (0,1.3) $) -- (2.7,1.3) |- (H0);
        \draw[line width=0.7pt,->,>=stealth,gray]
        (H0) -| (-2.7,1.3) -- ($ (C.west) + (0,1.3) $);

        \draw[line width=0.7pt,-,>=stealth,gray]
        ($ (C.east) + (0,1) $) -- (3,1) |- (H1);
        \draw[line width=0.7pt,->,>=stealth,gray]
        (H1) -| (-3,1) |- ($ (C.west) + (0,1) $);

        \draw[line width=0.7pt,-,>=stealth,gray]
        ($ (C.east) + (0,0.3) $) -- (3.6,0.3) |- (H2);
        \draw[line width=0.7pt,->,>=stealth,gray]
        (H2) -| (-3.6,0.3) |- ($ (C.west) + (0,0.3) $);

        \node at ($ (C.west) + (-0.35,0.75) $) {\scalebox{0.6}{$ \vdots $}};
        \node at ($ (C.east) + (0.25,0.75) $) {\scalebox{0.6}{$ \vdots $}};

        \fill[fill=orange!60]
        ($ (H0.north west) + (0,-0.1) $) --
        ($ (H0.south west) + (0,0.1) $) --
        ($ (H0.south west) + (0.1,0) $) --
        ($ (H0.south east) + (-0.1,0) $) --
        ($ (H0.south east) + (0,0.1) $) --
        ($ (H0.north east) + (0,-0.1) $) --
        ($ (H0.north east) + (-0.1,0) $) --
        ($ (H0.north west) + (0.1,0) $) --
        cycle;

        \fill[fill=orange!60]
        ($ (H1.north west) + (0,-0.1) $) --
        ($ (H1.south west) + (0,0.1) $) --
        ($ (H1.south west) + (0.1,0) $) --
        ($ (H1.south east) + (-0.1,0) $) --
        ($ (H1.south east) + (0,0.1) $) --
        ($ (H1.north east) + (0,-0.1) $) --
        ($ (H1.north east) + (-0.1,0) $) --
        ($ (H1.north west) + (0.1,0) $) --
        cycle;

        \fill[fill=orange!60]
        ($ (H2.north west) + (0,-0.1) $) --
        ($ (H2.south west) + (0,0.1) $) --
        ($ (H2.south west) + (0.1,0) $) --
        ($ (H2.south east) + (-0.1,0) $) --
        ($ (H2.south east) + (0,0.1) $) --
        ($ (H2.north east) + (0,-0.1) $) --
        ($ (H2.north east) + (-0.1,0) $) --
        ($ (H2.north west) + (0.1,0) $) --
        cycle;

      \end{tikzpicture}
    };
  \end{tikzpicture}
  \vspace{-1em}
  \caption{TSL System Architecture \mbox{\ }}
  \label{fig:architecture2}
\end{figure}

\noindent only classified according to their arity, i.e., the number
of other function terms, they are applied to, as well as by their
type: input, output, cell, function or predicate.

TSL specifications are evaluated on a synchronous system architecture
as shown in \cref{fig:architecture2}. The syntax utilizes a term based
notion, build from input streams~$ \name{i} \in \inames $, output
streams~\mbox{$ \name{o} \in \onames $}, memory
cells~$ \name{c} \in \cells $, and function and predicate literals
$ \name{f} \in \fnames $ and $ \name{p} \in \pnames $ with
$ \pnames \subseteq \fnames $, respectively. The purpose of cells is
to memorize data values that had been output to a cell at
time~$ t \in \dtime $ for providing them again as inputs at
time~$ t + 1 $. We differentiate between function
terms~$ \fterm \in \fterms $ and predicate
terms~$ \pterm \in \pterms $, build according to the following
grammar:
\begin{equation*}
  \begin{array}{rl}
  \fterm \ \;:= & \name{s}_{\name{i}} \sep \name{f}~\;\fterm^{0}~\;\fterm^{1}~\;\cdots~\;\fterm^{n-1} \\[0.3em]
  \pterm \ \;:= & \name{p}~\;\fterm^{0}~\;\fterm^{1}~\;\ldots~\;\fterm^{n-1}
  \end{array}
\end{equation*}
Here, $ \name{s}_{\name{i}} \in \inames \cup \cells $ is either an
input stream or a cell. In a TSL formula~$ \varphi $, 
function terms are then combined to updates, extended with predicate terms, Boolean
connectives, and temporal operators:
\begin{equation*}
  \varphi \; := \; \pterm \ \; | \ \; \upd{\name{s}_{\name{o}}}{\fterm} \ \; | \ \; \neg \varphi \ \; | \ \; \varphi \wedge \varphi \ \; | \ \; \LTLnext \varphi \ \; | \ \; \varphi \LTLuntil \varphi
\end{equation*}
where $ \name{s}_{\name{o}} \in \onames \cup \cells $ is either an output signal or a cell.

The semantics of a TSL formula~$ \varphi $ utilize a universally
quantified assignment
function~$ \assign{\cdot} \from \fnames \to \functions $, fixing an
implementation for each predicate and function literal, as well as
input streams~$ \iota $. We only give an intuitive description of the
semantics here. For a fully formal description the interested reader is
referred to \cite{tsl}. Intuitively, the semantics of TSL are
summarized as follows:
\begin{itemize}

\item \textbf{Predicate terms} evaluate to either \texttt{true} or
  \texttt{false}, by first selecting implementations for all function
  and predicate literals according to $ \assign{\cdot} $, and then
  applying them to inputs, as given by $ \iota $, and cells, using the
  stored value at the current time~$ t $. The content of a cell
  thereby is fixed iteratively, by selecting the past values piped
  into the cell over time. Cells are initialized using a special
  constant, provided as part of~$ \assign{\cdot} $.

\item \textbf{Function terms} evaluate similar to predicate
  terms, except that they evaluate to values of arbitrary type.

\item \textbf{Updates} are used to pipe the results of function term
  evaluations to output streams or cells. Therefore, updates, as they
  appear in a TSL formula, semantically are typed as Boolean
  expressions. In that sense, update expressions state
  that a specific flow is executed at a specific time, where an
  update evaluates to \texttt{true} if it is used and to
  \texttt{false}, otherwise. Outputs or cells only can
  receive a single update at any time.

\item The \textbf{Boolean operators} \textit{negation}~[$ \neg $] and
\textit{conjunction}~[$ \wedge $], and the \textbf{temporal operators}
\textit{next}~[$ \mathop{\LTLnext} $] and
\textit{until}~[$ \hspace{0.6pt}\mathop{\LTLuntil} \hspace{0.2pt}$]
have standard semantics and feature the default derived operators such as
\textit{release} [$ \varphi \LTLrelease \psi \equiv \neg
$(($\neg \psi$)$ \LTLuntil $($\neg \varphi$))$ $],
\textit{finally}~[$ \LTLfinally \varphi \equiv \emph{true} \LTLuntil
\varphi $],
\textit{always}~[$ \LTLglobally \varphi \equiv \emph{false}
\LTLrelease \varphi $], and the \textit{weak} version of
\textit{until}
[$ \varphi \LTLweakuntil \psi \equiv (\varphi \LTLuntil \psi) \vee
(\LTLglobally \varphi) $]. The precedence order of the listed operators matches the listed order, except that $ \LTLglobally $ and $ \LTLfinally $ have higher precedence than $ \LTLuntil $ and $ \LTLrelease $.
\end{itemize}
The synthesis problem of creating a control flow architecture~$ A $
that satisfies a TSL specification~$ \varphi $ is stated by
\begin{equation*}
  \exists A. \ \, \forall \iota. \ \, \forall \assign{\cdot}. \ \, \branch{A}{\iota}, \iota \sats \varphi
\end{equation*}
where $ \branch{A}{\iota} $ denotes the output produced by $ A $
under the \mbox{input}~$ \iota $. Note that $ A $ must satisfy the
specification for all possibly chosen function and predicate
implementations, as selected by~$ \assign{\cdot} $, and all possible
inputs~$ \iota $.

\section{Module Specifications}

\noindent We describe the development process of using TSL
for the creation of the \textit{Syntroids} game components. We discuss the full step-by-step design process for the LedMatrix module and highlight
some insights for the other modules.

\subsection{LedMatrix}\label{sec:ledMatrix}

\medskip

\begin{center}
  \scalebox{1.4}{\begin{tikzpicture}
    \clip (8.5,-5.4) rectangle (13.8,-3.2);
    
    \node {
      \usebox{\boxarchitecture}
    };
  \end{tikzpicture}}
\end{center}

\noindent The LedMatrix module is responsible for displaying images on
the physical LED matrix screen. Therefore, it has to send control data over
the hardware pins and needs to interact with a memory module serving
as video memory. It receives writing commands for individual pixels,
passed via \name{writecolor} and the coordinates \name{xcoordinate}
and \name{ycoordinate}, and a control bit~\name{write}, indicating
whether a pixel must be written to the video memory.  The module is
also able to provide the pixel's~\name{color}, determined by the delivered
coordinates (with a certain delay), which is, however,
not used by our application at the moment.

The LED matrix hardware interface splits the screen into two halves, which are operated
in parallel. On every half, the same single column is active at a time, while all other columns are 
turned off. The active column is selected by the 16-bit \mbox{$\name{coord}_y$-Pin}. Each half additionally uses a register for holding the content of the shown column and a
buffer register of the size of one row. By turning \name{driverPin} high
 the driver register can be turned off. If \name{bufferPin} is high, then the content of
 the buffer register is moved to the driver register.

 The writing procedure for operating the matrix cycles through all
 columns for writing to the buffer registers and for flushing the
 content to the driver register, before showing the corresponding
 column. The buffer registers are shift registers, which shift their
 content each time \name{extclock} rises (also referred to as clock)
 and hold \mbox{3-bit} color values outputted to $ \name{color}_{1} $
 and $ \name{color}_{2} $, one for each half, respectively.  Due to
 electrical characteristics of the LED matrix multiple LEDs may
 turn on, even if only a single LED is lighted up, if the data is
 written to fast. This effect is known as ghosting and is avoided by
 slowing down the writing process.

For writing to the video memory over \name{ramwrite}, a command consisting of a color and an address
needs to be transmitted. The video memory also delivers a pixel~\name{ramout}, if requested with an address over~\name{rampos}.
Writing and reading are independent, but it is only possible to read a single value at a time.

The specification of the LedMatrix module is given~by 
\mbox{$ \varphi_{\!A} \hspace{-1pt}\!\to\!\hspace{-1pt} ( \bigwedge_{i = 1}^2 \!\psi_i^I )\!  \land \globally ( \bigwedge_{i=1}^{21} \!\psi_i ) $} and covers the following tasks: 

\medskip

\subsubsection{Memory Interaction}
Whenever a color is taken from the video-memory, it must be preceded by a lookup action.
\begin{codeblockeq}
  \begin{array}{rcl}
    \psi_1 & \asgn & (\LTLnext \, \upd{\name{color}_{1}}{\name{ramout}}) \\ 
           && \quad \rightarrow  \upd{\name{rampos}}{\func{rampos}_{1}~\name{coord}_{x}~(\name{coord}_{y} + 1)} \\[0.4em]
    \psi_2 & \asgn & (\LTLnext \, \upd{\name{color}_{2}}{\name{ramout}}) \\ 
           && \quad \rightarrow \upd{\name{rampos}}{\func{rampos}_{2}~\name{coord}_{x}~(\name{coord}_{y} + 1)} \\[0.4em]
    \psi_3 & \asgn & (\LTLnext \, \upd{\name{color}_{R}}{\name{ramout}}) \\ 
           && \quad \rightarrow \upd{\name{rampos}}{\func{rampos}_{R}~\name{xcoordinate}~\name{ycoordinate}}
  \end{array}
\end{codeblockeq}
However, there is no lookup action initially.
\begin{codeblockeq}
  \begin{array}{rcl}  
  \psi_1^I & \asgn & \neg \upd{\name{color}_{1}}{\name{ramout}} \; \land \; \neg \upd{\name{color}_{2}}{\name{ramout}}\; \land \mbox{\,} \\
           & & \neg \upd{\name{color}_{R}}{\name{ramout}}
  \end{array}
\end{codeblockeq}
The literals $\name{coord}_x $ and $ \name{coord}_{y} $ are cells storing the \mbox{$ x $-~and} $ y $-coordinates internally. Note that we use notions such as $ (+ 1) $ for TSL literals without pre-assigned semantics for improved readability reasons.

Only if the write signal is high, the passed color~is written to the video memory. Otherwise it remains unchanged.
\begin{codeblock}
    \psi_4 & \asgn & \name{write} \to \upd{\name{ramwrite}}{\\ & & \quad \name{writeram}~\name{writecolor}~\name{xcoordinate}~\name{ycoordinate}} \\[0.4em]
    \psi_5 & \asgn & \neg \name{write} \to \upd{\name{ramwrite}}{\const{writeramnone}} 
\end{codeblock}
Finally, the color is output infinitely often.
\begin{codeblock}
  \psi_6 & \asgn & \finally \, \upd{\name{color}_{R}}{\name{ramout}}
\end{codeblock}
\vspace{-1em}
\subsubsection{External Clock Generation}
The clock is initially low and toggles between low and high infinitely often. We use~$ () $ after function literals to mark them as constants.
\begin{codeblock}
  \psi_2^I & \asgn & \upd{\name{extclock}}{\const{low}} \\[0.4em]
    \psi_7 & \asgn & \LTLfinally \, \upd{\name{extclock}}{\const{high}} \; \land \; \LTLfinally \, \upd{\name{extclock}}{\const{low}}
\end{codeblock}
Whenever the clock is high, then the outputs are stable.
\begin{codeblock}
    \psi_8 & \asgn & \upd{\name{extclock}}{\const{high}} \\ 
    & & \quad \to \upd{\name{color}_1}{\name{color}_1} \; \land \; \upd{\name{color}_2}{\name{color}_2} \\
    & & \quad \ \land \; \upd{\name{coord}_x}{\name{coord}_x} \; \land \; \upd{\name{coord}_y}{\name{coord}_y}
\end{codeblock}
\subsubsection{Pixel Updates}
The module changes the \mbox{$ x $-}co\-ordinate $ \name{coord}_{x} $ infinitely often
for printing out at every pixel.
\begin{codeblock}
    \psi_{9} & \asgn & \finally \, \upd{\name{coord}_x}{\name{coord}_x + 1}
\end{codeblock}
For writing the correct color between each generated clock cycle we output colors 
at both colors pins:
\begin{codeblock}
    \psi_{10} & \asgn & \upd{\name{extclock}}{\const{low}} \\
    & & \quad \to \big(\upd{\name{color}_1}{\name{ramout}} \release \neg \upd{\name{extclock}}{\const{high}}\big) \\[0.4em]
    \psi_{11} & \asgn& \upd{\name{extclock}}{\const{low}} \\
    & & \quad \to \big(\upd{\name{color}_2}{\name{ramout}} \release \neg \upd{\name{extclock}}{\const{high}}\big)
\end{codeblock}
However, both of these color settings are preceded by a ram lookup that requires the correct coordinates. 
Since the led matrix works by using a shift register, the module has to adjust the internal
$ x $-coordinate before looking up the colors, but only once for every clock cycle.
\begin{codeblock}
  \psi_{12} & \asgn & \upd{\name{extclock}}{\const{low}} \\
  && \quad \to \big(\upd{\name{coord}_x}{\name{coord}_x + 1} \release \neg ( \\
  && \qquad \quad \ \ \; \;\upd{\name{color}_1}{\name{ramout}} \; \lor \; \upd{\name{color}_2}{\name{ramout}}  \\
  && \qquad \quad \; \lor \; \upd{\name{rampos}}{\func{rampos1}~\name{coord}_x~(\name{coord}_y + 1)}\\
  && \qquad \quad \; \lor \; \upd{\name{rampos}}{\func{rampos2}~\name{coord}_x~(\name{coord}_y + 1)}\\
  && \qquad \quad \; \lor \; \upd{\name{extclock}}{\const{high}} )\!\big) \\[0.4em]
    \psi_{13} & \asgn & \upd{\name{coord}_x}{\name{coord}_x + 1} \to \LTLnext \big( \upd{\name{extclock}}{\const{high}}\\
    && \quad  \release \, \neg \upd{\name{coord}_x}{\name{coord}_x + 1} \big)
\end{codeblock}
When reaching the maximum $ x $-value the module has to adjust the $ y $-coordinate for writing the next row
\begin{codeblock}
  \psi_{14} & \asgn & \upd{\name{coord}_x}{\name{coord}_x + 1} \land (\name{coord}_x = \constidx{size}{x} - 1) \\
  && \quad \leftrightarrow \upd{\name{coord}_y}{\name{coord}_y + 1}
\end{codeblock}
while exactly then, it also prints the content of the buffer.
\begin{codeblock}
    \psi_{15} & \asgn & \upd{\name{bufferPin}}{\const{high}}\\ 
    &    & \quad \leftrightarrow \big( \LTLnext \, \upd{\name{coord}_x}{\name{coord}_x + 1} \\
    & & \qquad \ \ \;  \land \; (\name{coord}_x = \constidx{size}{x} - 1 )\big) \\[0.4em]
    \psi_{16} & \asgn & \upd{\name{bufferPin}}{\const{high}} \; \lor \; \upd{\name{bufferPin}}{\const{low}}
\end{codeblock}
Note that formula~$ \psi_{15} $ avoids that: if not changing the output, the
buffer pass-through is never active. As the driver pin is not used we fix it to be always low:
\begin{codeblock}
    \psi_{17} & := & \upd{\name{driverPin}}{\const{low}}
\end{codeblock}
\subsubsection{Ghosting Elimination}\label{subsec:ghosting}
The last subroutine introduces delay to avoid ghosting.
The delay itself is handled by \name{waitcounter}.
If the clock is low, then the counter starts. The $ x $-coordinate does not change 
until it reaches zero again due to an overflow.
\begin{codeblock}
    \psi_{18} & \asgn & \un{extclock}{\const{low}} \\
    & & \quad \to  \neg \upd{\name{waitcounter}}{\name{waitcounter} + 1} \; \land  \\
    & & \quad \phantom{to} \; \upd{\name{coord}_x}{\name{coord}_x + 1} \\[0.4em]
    \psi_{19} & \asgn & \neg (\func{eqz} \, \name{waitcounter}) \\ 
               & & \quad \to \upd{\name{waitcounter}}{\name{waitcounter} + 1} \\[0.4em]
    \psi_{20} & \asgn & \upd{\name{waitcounter}}{\name{waitcounter} + 1}\\
               & & \quad \to (\name{waitcounter} = 0) \lor \un{extclock}{\const{low}}
\end{codeblock}
However, until the overflow, $\name{coord}_{\name{x}}$ does not change
\begin{codeblock}
    \psi_{21} & \asgn & \un{extclock}{\const{low}} \to \LTLnext \big(  \neg \upd{\name{coord}_x}{\name{coord}_x + 1} \\
    &                &\quad  \; \until \; (\name{waitcounter} \neq 0)\big)
\end{codeblock}
and because of $\psi_{12}$ the whole writing process is stalled. 
The stalling must end eventually, which is satisfied since the counter overflows.
However, this behavior requires information on the data, i.e., the assumption
\vspace{0.5em}
\begin{codeblock}
    \varphi_{\!A} & \asgn & \globally \, \finally \, (\name{waitcounter} \neq 0) 
\end{codeblock}
\vspace{-1.2em}
\subsection{Scoreboard}\label{sec:scoreBoard}

\medskip

\begin{center}
  \scalebox{1.4}{\begin{tikzpicture}
    \clip (6,2.95) rectangle (11.5,4.6);
    
    \node {
      \usebox{\boxarchitecture}      
    };
  \end{tikzpicture}}
\end{center}

\noindent 
The Scoreboard module receives the \name{score}, its color \name{scorecolor}
and a Boolean signal~\name{gameover} indicating if the game is over. 
It provides a single point which consist of an $ x $-coordinate \name{bxcoord}, 
a \mbox{$ y $-coordinate} \name{bycoord} and a color value~\name{actcolor}, which may be written to the video-memory.
To draw the right image the module cycles over all pixels and writes the appropriate colors: 
\begin{codeblock}
    & \globally \, \un{xcoord}{\name{xcoord} + 1} \; \land &\\
    & \globally \, \big( (\name{xcoord} = \constidx{size}{x} - 1) \leftrightarrow \upd{\name{ycoord}}{\name{ycoord} + 1} \big) &
\end{codeblock}
where \name{xcoord} and \name{ycoord} are internal 5-bit values that
may overflow. Using such
counters is on the one hand necessary, since currently available
synthesis tools are not able to handle specifications with large
numbers of $\LTLnext$-chains, and on the other hand useful, as they automatically parameterize the module for possibly different screen sizes.

\vspace{0.5em}

\subsection{GameLogic}\label{sec:gameLogic}
\medskip

\begin{center}
  \scalebox{1.4}{\begin{tikzpicture}
    \clip (4.85,1.5) rectangle (10.7,3.6);
    
    \node {
      \usebox{\boxarchitecture}      
    };
  \end{tikzpicture}}
\end{center}

\noindent The GameLogic module manages the logic and state of the
game. It chooses between game over and running mode, selects the score,
and coordinates the enemies with the actions of the player.  Among
others, it receives Boolean signals indicating the \name{gamestart}
and whether the player has \name{shot}, as well as the data of all enemies
bundled together to \name{enemies}. Properties of individual enemies
are selected with functions like \name{getenemyangle} and
\name{getenemyradius} using the enemy index, realized through the
internal cell \name{counter}. A useful design feature of TSL is that
it automatically ensures conflict free management of streams, even at
different places. An example is the output
stream~\name{score}, which depends on the game state satisfying
\vspace{-0.8em}
\begin{codeblockeq}
    \globally \, \big(\name{gamestart} \leftrightarrow \un{score}{\const{zeroScore}} \big)
\end{codeblockeq}
The property states that the \name{score} is reset to zero if the
game restarts.
At the same time the condition
\begin{codeblockeq}
  \globally \, \big(\un{gameover}{\const{high}} \land \neg \name{gamestart} \to \un{score}{\name{score}}\big)
\end{codeblockeq}
requires that if the game is over and is not restarted, then the score
does not change, which ensures that the score is not changed when
the game is over. The semantics of TSL ensure that both properties
are realizable simultaneously, while the synthesis engine takes care
that there indeed is a conflict free resolution. The feature especially pays
off as soon as more properties are added.
For example, another requirement is the correct coordination of enemies and player actions. Especially, if the player shoots an enemy
the score must be increased.
\begin{codeblockeq}
  \begin{array}{l}
    \globally \LTLnext \big(\neg \name{gamestart} \land \un{gameover}{\const{low}} \\
    \quad \to \big(\un{score}{\name{score} + 1} \leftrightarrow     \big((\name{shotCounter} > 0) \; \land\\
    \qquad \qquad \func{hitenemy}~(\func{getenemyangle}~\name{enemies}~\name{counter})\\
    \qquad \qquad \quad (\func{getenemyangle}~\name{enemies} \name{counter})~\name{rotation} \big) \big) \big)
  \end{array}
\end{codeblockeq}

\vspace{-0.5em}

\subsection{Sensor}\label{sec:sensor}

\noindent The \textit{Sensor} module coordinates the different sensor submodules
via the \name{partControl} output selecting the signal from each
module.  It has to meet the following properties:
\begin{enumerate}
\item The sensor must be initialized first.
\item All reading submodules must be started repeatedly.
\item An active module has to finish before starting the next one.
\item It is forbidden to repeat the initialization.
\end{enumerate}
The second and third condition are declared as follows:
\begin{codeblockeq}
  \!\!\begin{array}{rcl}
    \psi_1 &\asgn&\LTLglobally \, \LTLfinally \, \un{partControl}{\const{accOn}}\\[0.4em]
    \psi_2 &\asgn&\LTLglobally \, \LTLfinally \, \un{partControl}{\const{gyrOn}}\\[0.4em]
    \psi_3 &\asgn& \LTLglobally\, (\lnot \un{partControl}{\const{initOn}} \, \land \\
    &&\phantom{\LTLglobally \, (} \neg\name{gyrFinished} \, \land \, \neg \name{accFinished} \, \land \, \neg\name{initFinished} \\
           &&\qquad \ \  \rightarrow \un{partControl}{\const{noCmd}} )
  \end{array}
\end{codeblockeq}
The update~$\un{partControl}{\const{initOn}}$ is necessary for the
specification to be realizable.  Otherwise, the initialization would
be forbidden, since there is no finished signal initially.
It is assumed that the other modules will return a \textit{finished} signal after being started
\begin{codeblock}
  \\[-2.5em]
    \varphi_1 & \asgn &\un{partControl}{\const{accOn}} \,\to \,  \LTLnext  \LTLfinally \, \name{accFinished}\\
    \varphi_2 & \asgn &\un{partControl}{\const{gyrOn}} \, \to \, \LTLnext  \LTLfinally \, \name{gyrFinished}
\end{codeblock}
which is necessary for realizability, since
otherwise the eventuality cannot be satisfied if all inputs are always low.

\subsection{SensorPart}\label{sec:sensorPart}
\noindent This module is configured using six register addresses, a
sensor type determining the right chip select and a module type to choose
the correct register.
In the specification the following structure is used repeatedly
\begin{codeblockeq}
    \varphi_{\name{A}} \LTLrelease \, ((\varphi_{\name{A}} \rightarrow \varphi_{\name{B}}) \land (\neg \varphi_{\name{A}} \rightarrow \varphi_{\name{C}}))
\end{codeblockeq}
where $ \varphi_{\name{A}} $ depends on an input and
$ \varphi_{\name{B}} $ and $ \varphi_{\name{C}} $ are output
assigning updates, the formula specifies a state in which the module
waits for $ \varphi_{\name{A}} $, meanwhile outputting
$ \varphi_{\name{C}} $. If $ \varphi_{\name{A}} $ happens, then it switches
the state with output $ \varphi_{\name{B}} $.  The formula is used to
ensure a sequence of actions. A followup state 
is defined using a $ \LTLnext $-operation and the same structure as in the specification
above.
\begin{codeblockeq}
  \varphi_{\name{B}} \rightarrow \, \LTLnext \, (\varphi_{\name{A}} \LTLrelease \, ((\varphi_{\name{A}} \rightarrow \varphi_{\name{D}}) \land (\neg \varphi_{\name{A}} \rightarrow \varphi_{\name{C}})))
\end{codeblockeq}
The formula structure is used to sequentially read all six registers, to finish and to wait for the next start signal.
We use \textsc{answer} as an alias for $\func{spiFinished}~\name{spiResponse}$.
\begin{codeblockeq}
\begin{array}{l}
  \LTLglobally \big( \upd{\name{spiControl}}{~\func{readCmd}~\name{\constidx{reg}{1}}} \rightarrow \LTLnext \big(\textsc{answer}  \\
  \qquad \LTLrelease  \big( \phantom{|\land\,} (\phantom{\neg}\textsc{answer} \rightarrow \upd{\name{spiControl}}{\func{readCmd}~\name{\constidx{reg}{2}}})\\
  \qquad \phantom{\LTLrelease \big(} \land (\neg\textsc{answer} \rightarrow \un{spiControl}{\const{noCmd}})\big)\big)\\
\end{array}
\end{codeblockeq}
The property is repeated for registers $\name{\constidx{reg}{2}},\dots,\name{\constidx{reg}{6}}$.
There is a \textit{RegManager} command generated after each reception of an answer,
which is specified by equivalence with the next read command, executed at the same time.
\begin{codeblockeq}
  \begin{array}{l}
    \\[-0.4em]
  \LTLglobally \, \upd{\name{spiControl}}{\func{readCmd}~\name{\constidx{reg}{1}}}\\
  \quad \leftrightarrow \upd{\name{regManagerCmd}}{\func{setRegister} \\ \qquad \quad \ \ \const{moduleType}~\const{zero}~\name{spiResponse}}  
\end{array}
\end{codeblockeq}

\tableData

\section{Experimental Results}

\noindent With specifications for all modules at hand, we first
synthesize the control using the TSL synthesis toolchain~\cite{tslfrp}
in combination with the game based LTL synthesizer
\textsc{Strix}~\cite{strix} and the bounded synthesizers
\textsc{BoSy}~\cite{bosy} and \textsc{BoWSer}~\cite{DBLP:journals/corr/abs-1904-07736}.
As a result, we obtain a source code module for every synthesized component that
is implemented for the hardware description language
{\clash}~\cite{clash2015}
and parameterized in the universally quantified functions. These
parameters then are instantiated with manually created implementations
for 42 functions, 24 predicates and 10 data types, implemented with
less than 200 lines of {\clash} code. The modules then are wired together according to \cref{fig:architecture} and 
compiled to Verilog code using the {\clash}
compiler. Finally, using the open synthesis framework
\textsc{Yosys} and the place-and-route tool
\textsc{Nextpnr}~\cite{yosys} the code is turned into a
binary to be uploaded to the FPGA.

The project is implemented on an 
\textit{icoBoard}
with an \textit{iCE40 hx8k FPGA} providing 7680~LCs and a 100MHz clock. Due to timing constraints, however, it runs on a prescaled clock of 10MHz.
The screen is an Adafruit 
LED matrix 
consisting of $ 32 \times 32 $ RGB LEDs. Input movements are obtained
from a Digilent
PModNav 
module featuring an accelerometer and a gyroscope sensor. All sources
of the project are available at:
\begin{center}
  \texttt{\url{react.uni-saarland.de/casestudies/syntroids}}
\end{center}

Our experimental results for synthesizing control flow architectures from TSL are depicted in \cref{table_results}.
For each module we counted the number of guarantees (\textbf{G}) and assumptions (\textbf{A})
split into temporal (\textbf{T}) and non-temporal (\textbf{L}) sub-formulas.
Whenever possible, we used
\textsc{BoWSer} to determine the number of states of the smallest Mealy machine satisfying 
the specification (\textbf{M}).
For each tool and module, we measured the synthesis time in seconds
(\textbf{Time}) and the number of
AIGER 
latches (\textbf{Lat}) and gates (\textbf{Gat}) of the generated
circuit,
where for each module the highlighted result was used in the final implementation.

\diagramall

We also compared the synthesized game with a manually created reference implementation with respect to
the number of logic cells (LCs) used and timing guarantees provided by \textsc{Nextpnr}.
The results of \cref{fig:diagramall} show differences in  
LCs and timing when swapping a module in the hand-made game with a synthesized one.
The used abbreviations are defined in \cref{table_results}.  Similar
results are shown by \cref{fig:diagram}, except that there a group of
modules is swapped. All modules being swapped is
reflected by \textit{All}.

\diagram

\vspace{-0.2em}

\section{Discussion}

\noindent Our study shows that TSL synthesis provides several advantages over manual programming.

\subsubsection{Behavior Descriptions}
One major advantage of synthesis is that a specification describes control behavior much better than a classic program or hardware description.
The following situations provide some examples:
\paragraph{Data Manipulation at different Places} 
If data is manipulated that depends on many different logical conditions, which might even be part of different sub-routines, then TSL outperforms classically created code. An example is the \name{score} value of the \nameref{sec:gameLogic} module, which is manipulated at different 
places and depends on many different conditions. When handling \name{score} by hand, e.g., in {\clash} or Verilog, the
value that is output to \name{score} must be handled consistently. Hence, it must
be guarded by the right conditions, which, however, are also affected at all positions, where \name{score} is currently used. This is not only tedious, but also a highly error-prone task.

\paragraph{Scheduling by Order Constraints}
Using a partial order for describing how events must follow each other
is much easier than always fixing a total order. Examples are 
the conditions described in the \nameref{sec:sensorPart} module of the form
$\varphi_{\name{B}} \rightarrow \LTLnext\, (\varphi_{\name{D}} \LTLrelease ((\varphi_{\name{D}} \rightarrow \varphi_{\name{E}}) \land (\neg \varphi_{\name{D}} \rightarrow \varphi_{\name{F}}))) $
or conditions that reference updates, which will happen in the future, as in the \nameref{sec:ledMatrix} module.
\begin{codeblockeq}
    \begin{array}{l}
    \LTLglobally \big( (\LTLnext \, \upd{\name{color}_{1}}{\name{ramout}}) \\ 
      \qquad \ \ \rightarrow  \upd{\name{rampos}}{\func{rampos}_{1}~\name{coord}_x~(\name{coord}_y + 1)} \big)\\[-1em]
\end{array} 
\end{codeblockeq}

\paragraph{Schedulability} 
Stating that updates happen repeatedly is easy to specify 
using $\LTLglobally \, \LTLfinally$, without the need of giving any fixed
order, e.g., in the \nameref{sec:sensor} module:
\begin{codeblock}
  \\[-0.4em]
    \psi_1 & \asgn &\LTLglobally\LTLfinally \un{partControl}{\const{accOn}}\\
    \psi_2 & \asgn &\LTLglobally\LTLfinally \un{partControl}{\const{gyrOn}}\\
\end{codeblock}

\subsubsection{Optimally Timed Solutions}
Another interesting observation is that synthesis tools are able to
create optimally timed solutions, e.g.\ if there is an update that
happens repeatedly, then it is possible to specify its length by using
a hard bound in form of a $ \LTLnext $-chain.  An example is a
simplified version of the \nameref{sec:ledMatrix} specification that
does not take care of ghosting. In this case, the internal
\mbox{$ x $-coordinate}  must be
increased infinitely often. The cycle length between these increases
can be specified using
$\LTLnext^n \upd{\name{coord}_x}{\name{coord}_x + 1} $ for
$n \in \mathbb{N}$. In a separate test
series we found, that with $n = 5$ the module is realizable, but with
$n = 4$ it is not. Therefore, the minimal cycle length is five.
Hence, it is easily possible to enforce the minimal ``time density''
of cyclic behaviour, which would be hard to provably achive in a manual
implementation.

\subsubsection{Easy expandability}
Another advantage is that modules can be easily expanded by adding new properties to the
specification. An example is the \nameref{subsec:ghosting}, which can be added to the 
\nameref{sec:ledMatrix} specification without the need of changing any of the remaining properties.

\subsubsection{Modification and Reuse}
Due to the separation of data and control, a module can be modified
through small changes on the data level, without affecting its
properties on the control level. For example the stalling time of the
\nameref{sec:ledMatrix} or the screen size handled by
\nameref{sec:scoreBoard} are easy to change on the data level.
Also modules can be reused for similar tasks, which differ only in the data they work on.
An example is the SensorPart module, which is used multiple times to implement different parts of the sensor by instantiating it with different function implementations on the data level.

\subsubsection{Verification}

Synthesis has the verification problem included. It is especially easy to add new conditions,
which would not be necessary for determining the behavior, but are important additional safety conditions.

\section{Further Work}

\noindent
There are are also several open challenges.

\subsubsection{Synthesis Times}
TSL synthesis is a fairly hard problem such that it was foreseeable that synthesis tools took quite some time for the more sophisticate specifications (cf.~\cref{table_results}). Thus, these specifications indicate benchmarks,  
for which synthesis tools still have to leverage improvements for the future.

\subsubsection{Next Chains}
Synthesis tools yet are not able to cope with long chains of
$\LTLnext$, e.g., when describing a bounded waiting process. Although
TSL allows to circumvent the problem by pushing waiting times to data
counters, it may be the intend of the developer to specifically
constraint the bounded behavior at the control level.

\subsubsection{Specification Debugging}
As specifications are still written by humans, and
humans are prone to make mistakes, the specifications still might be
incorrect, especially finding unrealizability reasons is difficult.
Hence, we need better debugging tools that help
with identifying the mistakes and provide strategies for their resolution.

\subsubsection{Module Distribution}
TSL synthesis allows the creation of modules independently of each
other, to be finally composed to a single architecture, like we did
with \cref{fig:architecture}. However, it might be necessary to
specify global properties of the system to ensure the correct
interaction of multiple modules as well. For example, a related problem, that we
encountered, was that using multiple specifications does not prevent the
 introduction of latch-free cycles on paths between multiple
modules. We had to introduce them manually (cf. unlabeled gray boxes
in \cref{fig:architecture}) while taking care that they indeed
preserve the intended behavior.

\section{Conclusion}

\noindent We have presented \textit{Syntroids}, the first interactive and
reactive hardware game that has been completely specified with
Temporal Stream Logic and is synthesized from the created
specifications using current state-of-the-art synthesis tools. Our
experience shows that Temporal Stream Logic is indeed a feasible
design flow for the development of reactive systems, providing significant advantages over manual programming.
We also identified challenges that remain
to be solved in order to accomplish a robust \mbox{TSL-based} development process.

\medskip

\noindent \textbf{Acknowledgements.} We thank Dan Gisselquist for
debugging a design issue with us that lead to nondeterministic timing
behavior and the \textsc{Yosys} and {\clash} development teams for
rapidly resolving our reported issues.

\bibliographystyle{IEEEtran}
\bibliography{biblio}

\begin{thebibliography}{1}
\providecommand{\url}[1]{#1}
\csname url@samestyle\endcsname
\providecommand{\newblock}{\relax}
\providecommand{\bibinfo}[2]{#2}
\providecommand{\BIBentrySTDinterwordspacing}{\spaceskip=0pt\relax}
\providecommand{\BIBentryALTinterwordstretchfactor}{4}
\providecommand{\BIBentryALTinterwordspacing}{\spaceskip=\fontdimen2\font plus
\BIBentryALTinterwordstretchfactor\fontdimen3\font minus
  \fontdimen4\font\relax}
\providecommand{\BIBforeignlanguage}[2]{{%
\expandafter\ifx\csname l@#1\endcsname\relax
\typeout{** WARNING: IEEEtran.bst: No hyphenation pattern has been}%
\typeout{** loaded for the language `#1'. Using the pattern for}%
\typeout{** the default language instead.}%
\else
\language=\csname l@#1\endcsname
\fi
#2}}
\providecommand{\BIBdecl}{\relax}
\BIBdecl

\bibitem{amba}
R.~{Bloem}, S.~{Galler}, B.~{Jobstmann}, N.~{Piterman}, A.~{Pnueli}, and
  M.~{Weiglhofer}, ``Automatic hardware synthesis from specifications: A case
  study,'' in \emph{2007 Design, Automation Test in Europe Conference
  Exhibition}, April 2007, pp. 1--6.

\bibitem{tsl}
\BIBentryALTinterwordspacing
B.~Finkbeiner, F.~Klein, R.~Piskac, and M.~Santolucito, ``Temporal stream
  logic: Synthesis beyond the bools,'' in \emph{Computer Aided Verification -
  31th International Conference, {CAV} 2019, New York, NY, USA, July 15-18,
  2019, Proceedings, Part {I}}, 2019. [Online]. Available:
  \url{https://doi.org/10.1007/978-3-030-25540-4_35}
\BIBentrySTDinterwordspacing

\bibitem{clash2015}
C.~Baaij, ``Digital circuit in c$\lambda$ash: functional specifications and
  type-directed synthesis,'' Ph.D. dissertation, 1 2015, eemcs-eprint-23939.

\bibitem{tslfrp}
B.~Finkbeiner, F.~Klein, R.~Piskac, and M.~Santolucito, ``Synthesizing
  functional reactive programs,'' \emph{CoRR}, vol. abs/nnnn.nnnnn, 2019,
  available at
  \url{https://www.react.uni-saarland.de/publications/FKPS19b.html}.

\bibitem{strix}
\BIBentryALTinterwordspacing
P.~J. Meyer, S.~Sickert, and M.~Luttenberger, ``Strix: Explicit reactive
  synthesis strikes back!'' in \emph{Computer Aided Verification - 30th
  International Conference, {CAV} 2018, Held as Part of the Federated Logic
  Conference, FloC 2018, Oxford, UK, July 14-17, 2018, Proceedings, Part {I}},
  ser. Lecture Notes in Computer Science, H.~Chockler and G.~Weissenbacher,
  Eds., vol. 10981.\hskip 1em plus 0.5em minus 0.4em\relax Springer, 2018, pp.
  578--586. [Online]. Available:
  \url{https://doi.org/10.1007/978-3-319-96145-3\_31}
\BIBentrySTDinterwordspacing

\bibitem{bosy}
\BIBentryALTinterwordspacing
P.~Faymonville, B.~Finkbeiner, and L.~Tentrup, ``Bosy: An experimentation
  framework for bounded synthesis,'' in \emph{Computer Aided Verification -
  29th International Conference, {CAV} 2017, Heidelberg, Germany, July 24-28,
  2017, Proceedings, Part {II}}, ser. Lecture Notes in Computer Science,
  R.~Majumdar and V.~Kuncak, Eds., vol. 10427.\hskip 1em plus 0.5em minus
  0.4em\relax Springer, 2017, pp. 325--332. [Online]. Available:
  \url{https://doi.org/10.1007/978-3-319-63390-9\_17}
\BIBentrySTDinterwordspacing

\bibitem{DBLP:journals/corr/abs-1904-07736}
\BIBentryALTinterwordspacing
S.~Jacobs, R.~Bloem, M.~Colange, P.~Faymonville, B.~Finkbeiner, A.~Khalimov,
  F.~Klein, M.~Luttenberger, P.~J. Meyer, T.~Michaud, M.~Sakr, S.~Sickert,
  L.~Tentrup, and A.~Walker, ``The 5th reactive synthesis competition
  {(SYNTCOMP} 2018): Benchmarks, participants {\&} results,'' \emph{CoRR}, vol.
  abs/1904.07736, 2019. [Online]. Available:
  \url{http://arxiv.org/abs/1904.07736}
\BIBentrySTDinterwordspacing

\bibitem{yosys}
\BIBentryALTinterwordspacing
D.~Shah, E.~Hung, C.~Wolf, S.~Bazanski, D.~Gisselquist, and M.~Milanovic,
  ``Yosys+nextpnr: an open source framework from verilog to bitstream for
  commercial fpgas,'' \emph{CoRR}, vol. abs/1903.10407, 2019. [Online].
  Available: \url{http://arxiv.org/abs/1903.10407}
\BIBentrySTDinterwordspacing

\end{thebibliography}

\end{document}